\documentclass[11pt]{article}
\usepackage{epsfig} 
\usepackage{graphics}
\newcommand{\sect}[1]{ \section{#1} \setcounter{equation}{0} } 

\newcommand{\half}{\mbox{\small{$\frac{1}{2}$}}}

\newcommand{\Nf}{N_{\!f}}
\newcommand{\NF}{N_{\!F}}

\newcommand{\pslash}{p \! \! \! /}
\newcommand{\MSbar}{\overline{\mbox{MS}}}
\newcommand{\MSbars}{\overline{\mbox{\footnotesize{MS}}}}
\newcommand{\RI}{\mbox{RI${}^\prime$}}
\newcommand{\RIs}{\mbox{\footnotesize{RI${}^\prime$}}}

\begin{document}

\title{Tensor current renormalization in the RI${}^\prime$ scheme at four 
loops}
\author{J.A. Gracey, \\ Theoretical Physics Division, \\ 
Department of Mathematical Sciences, \\ University of Liverpool, \\ P.O. Box 
147, \\ Liverpool, \\ L69 3BX, \\ United Kingdom.} 
\date{}

\maketitle 

\vspace{5cm} 
\noindent 
{\bf Abstract.} To assist the matching of lattice field theory results to the
high energy continuum limit we evaluate the Green's function where the tensor
quark bilinear operator is inserted at zero momentum in a quark $2$-point
function for an arbitrary covariant gauge. This is carried out in both the 
$\MSbar$ and $\RI$ schemes to four loops. The tensor current anomalous 
dimension is also calculated to four loops in both schemes for an arbitrary 
colour group.

\vspace{-15.5cm}
\hspace{13.2cm}
{\bf LTH 1312}

\newpage 

\sect{Introduction.}

The Standard Model (SM) is generally accepted to be the core theory that 
describes particle dynamics operating at the current energy scales of the Large
Hadron Collider (LHC). However, as data gathering increases over the next few 
years it is expected that discrepancies between experimental results and SM 
predictions will emerge. This is on top of the known difficulty in reconciling 
small neutrino mass observations with the present neutrino content of the SM. 
Another aspect of the Standard Model that is the subject of intense study
centres on determining the precise numerical elements of the
Cabibbo-Kobayashi-Maskawa (CKM) matrix which governs quark mixing and underlies
CP violation. Ensuring that the independent parameters are calculated
accurately theoretically based on the Standard Model is an important foundation
to finding a discrepancy with experimental measurements. Equally if differences
are discovered one question that arises is what extension to the SM will 
explain the new observations. In this respect one major activity centres on 
constructing effective field theories that use the present SM particle content
to build dimension five and six operators. These operators can have CP 
violating or other properties, for instance. One subset of such effective 
theories is to incorporate extra interactions with Lorentz structures different
from those already in the Standard Model. For example, the SM has vector-axial 
vector interactions. However at an early stage of the SM development an 
alternative structure was considered that involved a tensor current. This was 
eventually excluded by experiment. In seeking to explore beyond the SM at 
current LHC energies, however, tensor couplings have become an area of interest
again. For example tensor couplings have been used to examine $\beta$ decay of 
the nucleon in addition to beyond the SM CP violation searches. Its effect is 
manifest in the neutron electric dipole moment. Similar tensor couplings are 
also of interest in rare $B$ decays as well as being included in SM effective 
theories in order to provide the freedom to cover the parameter search space as
widely as possible ahead of more precise experimental data. Further background 
to theoretical aspects of these issues can be found, for example, in 
\cite{1,2,3}. Indeed the use of tensor couplings in effective field theory 
extensions of the SM have been discussed recently in \cite{4}.

One of the main theoretical tools used to determine precise values of the
various matrix elements that are central to these SM studies is lattice gauge
theory where the related Green's functions are calculated numerically to very 
high accuracy. However, as the underlying field theory of the strong sector, 
Quantum Chromodynamics (QCD), is regularized by discretizing continuous 
spacetime lattice results have to be extrapolated to the continuum limit by 
reducing the lattice spacing. Taking such a limit is not straightforward but to
assist with error analyses any numerical evaluation of a Green's function has 
to be consistent with its known high energy behaviour. In other words lattice 
results have to match onto the same Green's function but computed by contrast 
directly in the continuum perturbative theory. The latter can be deduced using 
standard methods to evaluate Feynman integrals but to ensure precision matching
and error analysis, having the matrix elements to as high a loop order as 
possible is important. There has been a large industry producing such 
perturbative calculations over the last few decades particularly in the case of
operator insertions in $2$-point functions. For instance, see 
\cite{5,6,7,8,9,10,11} for the early developments. In essence there are two 
classes of such Green's functions. One is where an operator is inserted at zero
momentum, \cite{5,6,7}. The other configuration is where the insertion is at 
non-zero momentum and additionally the momenta of the other external fields are
non-zero, \cite{8,9,10,11}. While the latter class corresponds to a 
non-exceptional momentum configuration it is technically more difficult to 
compute high loop order corrections in this instance. By contrast Green's 
functions with operators at zero momentum insertion in a $2$-point function 
effectively equate to a $2$-point rather than a $3$-point calculation and, 
moreover, can be determined to much higher orders perturbatively. 

This is the main aim of this article where given the importance of the tensor 
current in QCD for exploring beyond the Standard Model we will compute the 
Green's function with that operator inserted at zero momentum at four loops. 
This will extend the equivalent three loop exercise of $20$ years ago, 
\cite{7}, which has been used in various lattice analyses that are focused on
understanding the various SM extensions mentioned earlier such as $B$ meson 
decays, $\beta$ decay of nucleons, electic dipole moments of nucleons and the 
$J/\psi$ decay constant. See, for example, \cite{12,13,14,15,16,17,18} for 
several instances including the recent results of \cite{19,20}. Equally there 
are also applications of the tensor current to effective field theory 
formulations of the SM, \cite{21}. In particular we will compute the matrix 
element in the modified minimal subtraction ($\MSbar$) scheme which is the 
standard reference scheme for comparing to experiment. However, as lattice 
measurements are carried out in a lattice motivated renormalization scheme 
known as the modified Regularization Invariant ($\RI$) scheme, \cite{5,6}, we 
will also produce the Green's function in that scheme. Equally we will 
determine the anomalous dimension of the tensor operator to four loops in both 
schemes. We qualify this by noting that the $\MSbar$ four loop tensor anomalous
dimension is already available but only for the $SU(3)$ colour group, 
\cite{22}. We will provide the full four loop $\MSbar$ result for an arbitrary 
colour group. Although lattice computations of operator Green's functions are 
invariably performed in the Landau gauge we will take a more general point of 
view and carry out our calculations in an arbitrary linear covariant gauge.

The paper is organized as follows. We recall the basic field theory formalism
that allows us to compute the Green's function of interest with the tensor
operator insertion in Section $2$. This includes the renormalization aspects
and the definition of the $\RI$ scheme. The focus of Section $3$ is to record
explicit four loop expressions and in particular the value of the Green's
function as well as the $\MSbar$ operator anomalous dimension in both schemes.
Concluding remarks are provided in Section $4$ while an appendix records 
expressions for an arbitrary linear covariant gauge and colour group. 

\sect{Background.}

To begin with we define the Green's function that will be our focus which is
\begin{equation}
G^{\mu\nu}_{{\cal{O}}^T}(p) ~=~
\langle \psi(p) ~ [ \bar{\psi} \sigma^{\mu\nu} \psi ] (0) ~ \bar{\psi}(-p)
\rangle
\label{tensgf}
\end{equation}
where the tensor operator
\begin{equation}
{{\cal O}}^T ~=~ \bar{\psi} \sigma^{\mu\nu} \psi
\end{equation}
is inserted at zero momentum and the antisymmetric tensor $\sigma^{\mu\nu}$ is 
defined by
\begin{equation}
\sigma^{\mu\nu} ~=~ \half [ \gamma^\mu, \gamma^\nu ] 
\end{equation}
meaning that $G^{\mu\nu}_{{\cal{O}}^T}(p)$ is also antisymmetric. The first 
stage of determining the corrections to the Green's function is to decompose it
into its Lorentz components which are formally defined by
\begin{equation}
G^{\mu\nu}_{\cal{O}^T}(p) ~=~ \Sigma^{(1)}_{{\cal{O}}^T}(p) \sigma^{\mu\nu} ~+~ 
\Sigma^{(2)}_{{\cal{O}}^T}(p) \left(
\pslash \gamma^\mu p^ \nu ~-~ \pslash \gamma^\nu p^\mu \right) \frac{1}{p^2}
\end{equation}
where $\Sigma^{(i)}_{{\cal{O}}^T}(p)$ are scalar functions. These are isolated 
formally by a projection method, \cite{7}. Contracting the Green's function 
with two independent tensors produces a set of linear equations that can be 
inverted to deduce that
\begin{eqnarray}
\Sigma^{(1)}_{{\cal{O}}^T}(p) &=& -~ \frac{1}{4(d-1)(d-2)}
\left[ \mbox{tr} \left( \sigma_{\mu\nu} G^{\mu\nu}_{{\cal{O}}^T}(p) \right) ~+~
\frac{1}{p^2} \mbox{tr} \left( ( \pslash \gamma_\mu p_\nu
- \pslash \gamma_\nu p_\mu ) G^{\mu\nu}_{{\cal{O}}^T}(p) \right) \right] 
\nonumber \\
\Sigma^{(2)}_{{\cal{O}}^T}(p) &=& -~ \frac{1}{4(d-1)(d-2)}
\left[ \mbox{tr} \left( \sigma_{\mu\nu} G^{\mu\nu}_{{\cal{O}}^T}(p) \right) ~+~
\frac{d}{2p^2} 
\mbox{tr} \left( ( \pslash \gamma_\mu p_\nu - \pslash \gamma_\nu p_\mu ) 
G^{\mu\nu}_{{\cal{O}}^T}(p) \right) \right] \nonumber \\
\end{eqnarray}
where the trace $\mbox{tr}$ is over the spinor indices. This is carried out in 
$d$-dimensions since we will be regularizing dimensionally in 
$d$~$=$~$4$~$-$~$2\epsilon$ dimensions. At the outset we note that this is the
same set-up that was ultilized in \cite{7}. We have retained the same approach
to allow those interested in extending their lattice matching analyses to
readily adapt their programmes to include the new perturbative corrections.

We now summarize the technical aspects of evaluating (\ref{tensgf}). The main 
reason for writing the amplitudes as linear combinations of two projections is 
to evaluate them by automatic Feynman graph integration packages. In \cite{7} 
the three loop computations were carried out with the {\sc Mincer} package that
was originally devised and implemented in {\sc Schoonschip} in \cite{23} but 
recoded, \cite{24}, in the symbolic manipulation language {\sc Form} in 
\cite{25,26}. The {\sc Mincer} algorithm evaluates massless $2$-point Feynman 
integrals in $d$~$=$~$4$~$-$~$2\epsilon$ dimensions to three loops. Therefore 
as the operator ${\cal{O}}^T$ is inserted at zero momentum then the package 
could be used to determine $G^{\mu\nu}_{{\cal{O}}^T}(p)$ to three loops. To 
extend the results of \cite{7} to the next order we follow the same approach 
but use the more recent {\sc Forcer} package, \cite{27,28}. This is the 
successor to {\sc Mincer} in that it equally determines the $\epsilon$ 
expansion of $2$-point massless Feynman integrals in $d$-dimensions but at four
loops. In order to effect the evaluation of (\ref{tensgf}) the first stage is 
to generate the Feynman graphs which is achieved by using the {\sc Qgraf} 
package, \cite{29}. In addition to the $1$, $13$ and $244$ graphs at 
respectively one, two and three loops, there are $5728$ to determine at four 
loops. Once the graphs are generated the Lorentz and colour indices are 
appended to the fields and the QCD Feynman rules implemented. To handle the 
group theory we used the {\tt color.h} {\sc Form} module, based on the article 
\cite{30}, as it is designed to account for any new higher rank colour Casimirs
that will arise. This routine is applied to each graph in the automatic 
evaluation prior to applying the {\sc Forcer} component as it can be the case 
that for certain graphs the group factor turns out to be zero. After the group 
factor is determined and the $\epsilon$ expansion found the expressions for 
each graph are added. To ensure that a finite value (\ref{tensgf}) is returned 
we carry out an automatic renormalization using the method of \cite{31}. The 
key point is that the value of each graph is written as a function of the bare 
parameters. These are the gauge coupling constant $g$ and the covariant gauge 
parameter $\alpha$ where the Landau gauge corresponds to a value of zero. We 
have chosen to include $\alpha$ since the operator anomalous dimension is 
independent of the parameter in the $\MSbar$ scheme, \cite{38,39}, and this
property can be exploited for checking purposes. However this property does not
persist in other schemes such as $\RI$. To introduce counterterms to render 
(\ref{tensgf}) finite the bare parameters are replaced by their renormalized 
partners. As the coupling and gauge parameter first occur in the one loop graph
their renormalization constants are only required at three loops. The overall 
divergence that remains in the Green's function after completing this process 
is then absorbed by the operator renormalization constant at four loops in 
whichever renormalization scheme is required. 

There is one caveat to this in that the four loop quark wave function 
renormalization constant has to be included due to the external legs of 
(\ref{tensgf}). This and the other field wave function renormalization
constants are now available to {\em five} loops in the $\MSbar$ scheme. For the
extension of the three loop RI${}^\prime$ tensor operator anomalous dimension
we also need the quark wave function renormalization in that scheme but now at
four loops. Therefore we have separately renormalized the quark $2$-point
function at four loops using {\sc Forcer} which involved the evaluation of
$1422$ graphs at that order. We obtained the same four loop $\MSbar$ expression
for the quark anomalous dimension $\gamma_\psi(a,\alpha)$ as \cite{32,33,34}
where $a$~$=$~$g^2/(16\pi^2)$. However it is a straightforward exercise to 
determine the anomalous dimension in the $\RI$ scheme which is defined by the 
criterion given in \cite{5,6} which is
\begin{equation}
\left. \lim_{\epsilon \rightarrow 0} \left[
Z^{\RIs}_\psi \Sigma_\psi(p) \right] \right|_{p^2 \, = \, \mu^2} ~=~ \pslash
\label{zpsiri}
\end{equation}
at the subtraction point where $\Sigma_\psi(p)$ is the quark $2$-point 
function and $\mu$ is a mass scale necessary in the regularization to ensure
the coupling constant remains dimensionless in $d$-dimensions. In other words 
at the subtraction point there are no $O(a)$ corrections to $\Sigma_\psi(p)$. 
With this definition we verified the three loop $\RI$ expression of \cite{7,35}
for $\gamma_\psi(a,\alpha)$ and deduced
\begin{eqnarray}
\left. \frac{}{} \gamma_\psi^{\RIs}(a,0) \right|^{SU(3)} &=&
-~ [4 \Nf - 67] \frac{a^2}{3}   
+ [416 \Nf^2 + 1728 \zeta_3 \Nf - 17888 \Nf - 32778 \zeta_3 + 156963] 
\frac{a^3}{108} 
\nonumber \\
&& 
-~ [ 16000 \Nf^3 + 103680 \zeta_3 \Nf^2 - 1205680 \Nf^2 - 5834784 \zeta_3 \Nf 
+ 1075680 \zeta_5 \Nf
\nonumber \\
&& ~~~~
+ 24606080 \Nf + 62524516 \zeta_3 - 15846715 \zeta_5 
- 143460448 ] \frac{a^4}{1296} \nonumber \\
&& +~ O(a^5)
\end{eqnarray}
at four loops for $SU(3)$ or
\begin{eqnarray}
\left. \frac{}{} \gamma_\psi^{\RIs}(a,0) \right|^{SU(3)} &=&
-~ [ 1.333333 \Nf - 22.333333 ] a^2 \nonumber \\
&&
+~ [ 3.851852 \Nf^2 - 146.396719 \Nf + 1088.536841 ] a^3 \nonumber \\
&&
-~ [ 12.345679 \Nf^3 - 834.144090 \Nf^2 + 14434.984616 \Nf - 65381.420167 ] a^4
\nonumber \\
&& +~ O(a^5) 
\end{eqnarray}
numerically.

\sect{Tensor current results.}

Having described the computational technicalities we now record the results
relating to (\ref{tensgf}). First we recall the result of \cite{22} for the 
renormalization of the tensor current in the $\MSbar$ scheme for $SU(3)$ is
\begin{eqnarray}
\left. \frac{}{} \gamma_T^{\MSbars}(a) \right|^{SU(3)} 
&=& \frac{4}{3} a ~+~ \frac{2}{27} \left[ 543 - 26 \Nf \right] a^2      
\nonumber \\
&& +~ \left[ 52555 - 36 \Nf^2 - 1440 \zeta_3 \Nf - 5240 \Nf - 2784 \zeta_3 
\right] \frac{a^3}{81}    
\nonumber \\
&& +~ \left[ 
1152 \zeta_3  \Nf^3
+ 168 \Nf^3 
+ 66240 \zeta_3 \Nf^2 
- 25920 \zeta_4 \Nf^2 
+ 39844 \Nf^2 
\right. \nonumber \\
&& \left. ~~~~
- 1821984 \zeta_3 \Nf 
+ 377568 \zeta_4 \Nf
+ 993600 \zeta_5 \Nf 
- 3074758 \Nf 
\right. \nonumber \\
&& \left. ~~~~
- 742368 \zeta_3 
+ 826848 \zeta_4 
- 4018560 \zeta_5 
+ 19876653
\right] \frac{a^4}{1458} ~+~ O(a^5) ~~~~
\end{eqnarray} 
We have verified this as a corollary of the expression for an arbitrary colour
group which is 
\begin{eqnarray}
\gamma_T^{\MSbars}(a) &=& C_F a ~+~ 
\left[ 
\frac{257}{18} C_F C_A 
- \frac{26}{9} C_F T_F \Nf
- \frac{19}{2} C_F^2
\right] a^2
\nonumber \\
&& +~ \left[ 
\left[
\frac{13639}{108}
- 40 \zeta_3
\right]
C_F C_A^2
- \frac{4}{3} C_F T_F^2 \Nf^2
- \left[
\frac{1004}{27}
+ 16 \zeta_3
\right]
C_F C_A T_F \Nf
\right. \nonumber \\
&& \left. ~~~~~
+ \left[
\frac{98}{9}
+ 16 \zeta_3
\right]
C_F^2 T_F \Nf
+ \left[
112 \zeta_3
- \frac{6823}{36}
\right]
C_F^2 C_A
+ \left[
\frac{365}{6}
- 64 \zeta_3
\right]
C_F^3
\right] a^3
\nonumber \\
&& +~ \left[ 
\left[
\frac{208}{3} \zeta_3
- \frac{32}{3}
- \frac{640}{3} \zeta_5
\right]
\frac{d_F^{abcd} d_A^{abcd}}{\NF}
+ \left[
128
- 32 \zeta_3
\right]
\frac{d_A^{abcd} d_A^{abcd}}{\NF}
\right. \nonumber \\
&& \left. ~~~~~
+ \left[
\frac{56}{81}
+ \frac{128}{27} \zeta_3
\right]
C_F T_F^3 \Nf^3
+ \left[
\frac{194}{81}
- 32 \zeta_4
+ \frac{736}{9} \zeta_3
\right]
C_F C_A T_F^2 \Nf^2
\right. \nonumber \\
&& \left. ~~~~~
+ \left[
8 \zeta_4
- \frac{73409}{162}
+ \frac{400}{3} \zeta_5
- \frac{980}{3} \zeta_3
\right]
C_F C_A^2 T_F \Nf
\right. \nonumber \\
&& \left. ~~~~~
+ \left[
\frac{710581}{648}
- \frac{1600}{9} \zeta_5
+ 220 \zeta_4
- \frac{12598}{27} \zeta_3
\right]
C_F C_A^3
\right. \nonumber \\
&& \left. ~~~~~
+ \left[
\frac{4544}{81}
+ 32 \zeta_4
- \frac{736}{9} \zeta_3
\right]
C_F^2 T_F^2 \Nf^2
\right. \nonumber \\
&& \left. ~~~~~
+ \left[
\frac{523}{3}
+ \frac{80}{3} \zeta_5
+ 136 \zeta_4
+ \frac{2416}{9} \zeta_3
\right]
C_F^2 C_A T_F \Nf
\right. \nonumber \\
&& \left. ~~~~~
+ \left[
\frac{2320}{3} \zeta_5
- 616 \zeta_4
+ \frac{9800}{9} \zeta_3
- \frac{733979}{324}
\right]
C_F^2 C_A^2
\right. \nonumber \\
&& \left. ~~~~~
+ \left[
\frac{2900}{27}
- 160 \zeta_5
- 128 \zeta_4
- \frac{8}{9} \zeta_3
\right]
C_F^3 T_F \Nf
\right. \nonumber \\
&& \left. ~~~~~
+ \left[
\frac{179363}{108}
- \frac{4880}{3} \zeta_5
+ 352 \zeta_4
+ \frac{1012}{9} \zeta_3
\right]
C_F^3 C_A
\right. \nonumber \\
&& \left. ~~~~~
+ \left[
\frac{3200}{3} \zeta_5
- \frac{2000}{3} \zeta_3
- \frac{10489}{24}
\right]
C_F^4
\right] a^4 ~+~ O(a^5)
\label{gammaTms}
\end{eqnarray}
where $\zeta_n$ is the Riemann zeta function and $C_F$, $C_A$ and $T_F$ are
the standard colour factors. At four loops the rank $4$ colour Casimirs
$d_R^{abcd}$ arise with the tensor being defined by
\begin{equation}
d_R^{abcd} ~=~ \frac{1}{6} \mbox{Tr} \left( T^a T^{(b} T^c T^{d)}
\right)
\end{equation}
for the representation $R$ where the trace $\mbox{Tr}$ is over the colour 
indices of the matrices representing the group generators $T^a$. En route we 
have verified the earlier respective two and three loop terms derived in 
\cite{36,37}. For practical purposes we recall the numerical value is,
\cite{22},
\begin{eqnarray}
\left. \frac{}{} \gamma_T^{\MSbars}(a) \right|^{SU(3)} 
&=& 1.333333 a + \left[ 40.222222 - 1.925926 \Nf \right] a^2
\nonumber \\
&& +~ \left[ 607.512019 - 86.061258 \Nf - 0.444444 \Nf^2 \right] a^3 
\nonumber \\
&& +~ \left[ 1.065000 \Nf^3 + 62.698512 \Nf^2 - 2624.104532 \Nf 
+ 10776.573952 \right] a^4 \nonumber \\
&& +~ O(a^5)
\end{eqnarray}
for $SU(3)$.

While the $SU(3)$ value of $\gamma_T^{\MSbars}(a)$ was already available,
\cite{22}, what is one of the main results here is the extension of
$\Sigma^{(1)}_{{\cal{O}}^T}(p)$ to four loops. In particular we have  
\begin{eqnarray}
\left. \Sigma^{(1) ~ \MSbars}_{{\cal{O}}^T}(p)
\right|_{\alpha = 0}^{SU(3)} &=& 1 
+ \left[ 
\frac{76}{9} \zeta_3
- \frac{1693}{54}
+ \frac{124}{81} \Nf 
\right] a^2 
\nonumber \\
&&
+ \left[ 
\frac{22952}{243} \zeta_3
- \frac{277}{108} \zeta_4
- \frac{265}{81} \zeta_5
- \frac{1977125}{2916}
\right. \nonumber \\
&& \left. ~~~
+ \left[
\frac{63764}{729}
+ \frac{80}{9} \zeta_4
- \frac{776}{27} \zeta_3
\right] \Nf 
+ \left[
\frac{376}{2187}
- \frac{32}{81} \zeta_3
\right] \Nf^2 \right] a^3
\nonumber \\
&&
+ \left[ 
\frac{42157925}{31104} \zeta_6
- \frac{629370181}{23328}
- \frac{476917595}{15552} \zeta_7
+ \frac{2784917789}{46656} \zeta_5
\right. \nonumber \\
&& \left. ~~~
- \frac{1202905}{2592} \zeta_4
- \frac{538028059}{23328} \zeta_3
+ \frac{48310147}{15552} \zeta_3^2
\right. \nonumber \\
&& \left. ~~~
+ \left[
\frac{124447867}{17496}
+ \frac{2989}{2} \zeta_7
- \frac{12400}{27} \zeta_6
- \frac{5255677}{972} \zeta_5
+ \frac{283045}{648} \zeta_4
\right. \right. \nonumber \\
&& \left. \left. ~~~~~~~
+ \frac{1452433}{972} \zeta_3
- \frac{1880}{9} \zeta_3^2
\right] \Nf
\right. \nonumber \\
&& \left. ~~~
+ \left[
\frac{3320}{27} \zeta_5
- \frac{280}{27} \zeta_4
- \frac{2548}{81} \zeta_3
- \frac{13603319}{52488}
\right] \Nf^2
\right. \nonumber \\
&& \left. ~~~
+ \left[
\frac{4610}{6561}
- \frac{8}{27} \zeta_4
+ \frac{32}{243} \zeta_3
\right] \Nf^3 \right] a^4 ~+~ O(a^5) 
\end{eqnarray} 
in the Landau gauge at the subtraction point and
\begin{eqnarray} 
\left. \Sigma^{(2) ~ \MSbars}_{{\cal{O}}^T}(p)
\right|_{\alpha = 0}^{SU(3)} &=& 0
\end{eqnarray} 
for $SU(3)$ with
\begin{eqnarray}
\left. \Sigma^{(1) ~ \MSbars}_{{\cal{O}}^T}(p)
\right|_{\alpha = 0}^{SU(3)} &=& 
1 + \left[ 1.530864 \Nf - 0.277778 \alpha^2 - 2.424683 \alpha 
- 21.201149 \right] a^2 \nonumber \\
&& +~ \left[ 62.540409 \Nf - 1.046804 \alpha^3 - 17.386077 \alpha^2 
- 5.0434846 \alpha \Nf 
\right. \nonumber \\
&& \left. ~~~~~
- 10.680631 \alpha - 0.302962 \Nf^2 - 570.657293 \right] a^3
\nonumber \\
&& +~ \left[ 11.9280686 \alpha^2 \Nf - 9.031353 \alpha^4 - 118.197512 \alpha^3 
- 527.617841 \alpha^2 
\right. \nonumber \\
&& \left. ~~~~~
- 3.840997 \alpha \Nf^2 + 155.224548 \alpha \Nf - 1864.356282 \alpha 
+ 0.540244 \Nf^3 
\right. \nonumber \\
&& \left. ~~~~~
- 180.703312 \Nf^2 + 4513.065131 \Nf - 18365.189753 \right] a^4 \nonumber \\
&& +~ O(a^5)
\end{eqnarray} 
numerically for the non-zero amplitude. We note as was observed before,
\cite{7,22}, the value of the channel $2$ amplitude is zero at four loops not 
only for $SU(3)$ but also for a general colour group. This may in fact be true 
to all orders as a consequence of some symmetry restriction. We have recorded 
the full four loop expression for $\Sigma^{(1) ~ \MSbars}_{{\cal{O}}^T}(p)$ in 
Appendix A and provided its electronic representation in the associated data 
file together with other results relating to the tensor operator. To gauge the 
effect of the new correction when $\Nf$~$=$~$3$ we have 
\begin{eqnarray}
\left. \Sigma^{(1) ~ \MSbars}_{{\cal{O}}^T}(p)
\right|_{\alpha = 0}^{SU(3)\,\Nf=3} &=& 
1 \,-\, 16.608556 a^2 \,-\, 385.762720 a^3 \,-\, 6437.737582 a^4 \,+\, 
O(a^5) ~~~~~~~
\label{amp1ms}
\end{eqnarray} 
and with $\alpha_s$~$=$~$0.12$ its two, three and four loop values are
$0.998486$, $0.998150$ and $0.998096$ respectively using naive substitution.
Viewed this way one would imagine that the imperceptible difference between the
three and four loop results could lead to a minor refinement of the error on 
the lattice extrapolation to the high energy expression. This observation is 
one of the main consequences of our next order study. At this point we note 
that we found a discrepancy in the $\Nf$ independent part of the three loop 
term of $\Sigma^{(1) ~ \MSbars}_{{\cal{O}}^T}(p)$ given in \cite{7}. In 
particular it has only a minor effect in the $O(a^3)$ coefficient. For example 
evaluating the corresponding $O(a^3)$ coefficient of (\ref{amp1ms}) in \cite{7}
at $\Nf$~$=$~$3$ in the Landau gauge would have given $-$~$399.155300$. With 
this value then at three loops (\ref{amp1ms}) evaluates to $0.998084$ at
$\alpha_s$~$=$~$0.12$ so that there is no large discrepancy.

The focus so far has been on the $\MSbar$ scheme but in \cite{7} the operator
was also renormalized in the $\RI$ scheme. For completeness we also extend the 
three loop results for that scheme here. First we recall the definition of the
operator renormalization constant of \cite{7}, that has parallels with the
quark wave function renormalization definition of (\ref{zpsiri}), which is
\begin{equation}
\left. \lim_{\epsilon \, \rightarrow \, 0} \left[ Z^{\RIs}_\psi
Z^{\RIs}_{{\cal O}^T} 
\Sigma^{(1)}_{{\cal O}^T}(p) \right] \right|_{p^2 \, = \, \mu^2} ~=~ 1 ~.
\label{ztensri}
\end{equation} 
In other words the channel $1$ amplitude is used since the divergences that 
lead to the $\MSbar$ scheme renormalization are located there irrespective of
the fact that $\Sigma^{(2)}_{{\cal O}^T}(p)$ vanishes. With (\ref{ztensri})
the anomalous dimension $\RI$ scheme anomalous dimension 
$\gamma^{\mbox{\footnotesize{RI$^\prime$}}}_{{\cal{O}}^T}(a,\alpha)$ is
\begin{eqnarray}
\gamma^{\mbox{\footnotesize{RI$^\prime$}}}_{{\cal{O}}^T}(a,0) &=&
C_F a + C_F \left[ 257 C_A - 171 C_F - 52 \Nf T_F \right] \frac{C_A a^2}{18} 
\nonumber \\
&& +~ \left[ 
53387 C_A^2 
- 23112 \zeta_3 C_A^2
+ 41904 \zeta_3 C_A C_F
- 57186 C_A C_F 
+ 3456 \zeta_3 C_A \Nf T_F
\right. \nonumber \\
&& \left. ~~~~
- 24884 C_A \Nf T_F 
- 10368 \zeta_3 C_F^2
+ 9855 C_F^2 
- 6048 \zeta_3 C_F \Nf T_F
\right. \nonumber \\
&& \left. ~~~~
+ 11394 C_F \Nf T_F 
+ 2288 \Nf^2 T_F^2 \right] \frac{C_F a^3}{162} \nonumber \\
&& +~ \left[ 
97637317 C_A^3 C_F \NF 
+ 5196960 \zeta_5 C_A^3 C_F \NF
- 57962790 \zeta_3 C_A^3 C_F \NF
\right. \nonumber \\
&& \left. ~~~~
+ 103267872 \zeta_3 C_A^2 C_F^2 \NF
+ 1321920 \zeta_5 C_A^2 C_F^2 \NF
- 135883278 C_A^2 C_F^2 \NF 
\right. \nonumber \\
&& \left. ~~~~
+ 26573832 \zeta_3 C_A^2 C_F \NF \Nf T_F
- 1088640 \zeta_5 C_A^2 C_F \NF \Nf T_F
\right. \nonumber \\
&& \left. ~~~~
- 72145932 C_A^2 C_F \NF \Nf T_F 
- 20785248 \zeta_3 C_A C_F^3 \NF
\right. \nonumber \\
&& \left. ~~~~
- 17262720 \zeta_5 C_A C_F^3 \NF
+ 43519680 C_A C_F^3 \NF 
- 42000768 \zeta_3 C_A C_F^2 \NF \Nf T_F
\right. \nonumber \\
&& \left. ~~~~
+ 3110400 \zeta_5 C_A C_F^2 \NF \Nf T_F
+ 57759192 C_A C_F^2 \NF \Nf T_F 
\right. \nonumber \\
&& \left. ~~~~
- 2695680 \zeta_3 C_A C_F \NF \Nf^2 T_F^2
+ 15287808 C_A C_F \NF \Nf^2 T_F^2 
- 7776000 \zeta_3 C_F^4 \NF
\right. \nonumber \\
&& \left. ~~~~
+ 12441600 \zeta_5 C_F^4 \NF
- 5097654 C_F^4 \NF 
+ 6158592 \zeta_3 C_F^3 \NF \Nf T_F
\right. \nonumber \\
&& \left. ~~~~
- 2488320 \zeta_5 C_F^3 \NF \Nf T_F
- 5448384 C_F^3 \NF \Nf T_F 
+ 3525120 \zeta_3 \zeta_5 C_F^2 \NF \Nf^2 T_F^2
\right. \nonumber \\
&& \left. ~~~~
- 5053824 C_F^2 \NF \Nf^2 T_F^2 
- 872192 C_F \NF \Nf^3 T_F^3 
+ 808704 \zeta_3 d_F^{abcd} d_A^{abcd}
\right. \nonumber \\
&& \left. ~~~~
- 2488320 \zeta_5 d_F^{abcd} d_A^{abcd}
- 124416 d_F^{abcd} d_A^{abcd} 
- 373248 \zeta_3 d_F^{abcd} d_F^{abcd} \Nf
\right. \nonumber \\
&& \left. ~~~~
+ 1492992 d_F^{abcd} d_F^{abcd} \Nf \right] \frac{a^4}{11664 \NF} ~+~ O(a^5)
\label{gammaTri}
\end{eqnarray}
in the Landau gauge. Although the tensor operator is gauge invariant its
anomalous dimension will be dependent on the gauge parameter in general. It is 
only in the $\MSbar$ scheme that the anomalous dimension of a gauge invariant 
operator is independent of the gauge parameter, \cite{38,39}. Indeed that was a
check on the emergence of an $\alpha$ independent $\MSbar$ expression at four 
loops for a general colour group. For comparison we note
\begin{eqnarray}
\left. \frac{}{} \gamma^{\mbox{\footnotesize{RI$^\prime$}}}_{{\cal{O}}^T}(a,0)
\right|^{SU(3)} &=& 1.333333 a + [ 40.222222 - 1.925926 \Nf ] a^2
\nonumber \\
&& +~ [ 4.707819 \Nf^2 - 233.294078 \Nf + 1634.149833 ] a^3 \nonumber \\
&& +~ [ 88297.353564 - 18912.306371 \Nf + 1001.765247 \Nf^2 
\nonumber \\
&& ~~~~
-~ 12.462734 \Nf^3 ] a^4 ~+~ O(a^5) 
\end{eqnarray}
numerically. We do not need to record the $\RI$ expression for 
$\Sigma^{(1)}_{{\cal O}^T}(p)$ as trivially it will be unity by construction.

One check on our four loop $\RI$ tensor operator anomalous dimension is to 
exploit a useful property of the renormalization group formalism. If an
anomalous dimension is known at $L$ loops in one renormalization scheme and
has been renormalized to $(L-1)$ loops in another scheme one can deduce the
$L$ loop anomalous dimension in the latter scheme by using a conversion
function. This is defined as the ratio of the renormalization constants in the 
respective schemes at $(L-1)$ loops. In our particular case the conversion 
function $C_{{\cal O}^T}(a,\alpha)$ is defined by 
\begin{equation}
C_{{\cal O}^T}(a,\alpha) ~=~
\frac{Z^{\RIs}_{{\cal O}^T}}{Z^{\mbox{\footnotesize{$\MSbar$}}}_{{\cal O}^T}}
\label{confundef}
\end{equation}
where the variables of the argument are in the $\MSbar$ scheme. We note this
since $Z^{\RIs}_{{\cal O}^T}$ is a function of $a^{\RIs}$ and $\alpha^{\RIs}$
which would suggest that $C_{{\cal O}^T}(a,\alpha)$ is not dependent on
$\epsilon$ but also has poles in the regularization. This is not the case since
$a^{\RIs}$ and $\alpha^{\RIs}$ are functions of $a^{\MSbars}$ and 
$\alpha^{\MSbars}$ with the relation between the two sets being established to
three loops in \cite{7}. Strictly only the relation of the gauge parameters is
needed to this order since the coupling constants are the same in both schemes.
In \cite{7} the three loop terms of the gauge parameter map were actually
superfluous for the three loop check analogous to the one we will repeat here
but are needed at four loops. Once $C_{{\cal O}^T}(a,\alpha)$ is available the
anomalous dimensions between the two schemes are connected by
\begin{eqnarray}
\gamma^{\RIs}_{{\cal O}^T} \left(a_{\RIs}, \alpha_{\RIs}\right) &=&
\left[ \gamma^{\MSbars}_{{\cal O}^T} \left(a_{\MSbars}\right) ~-~
\beta\left(a_{\MSbars}\right) \frac{\partial ~~~}{\partial a_{\MSbars}}
\ln C_{{\cal O}^T}\left(a_{\MSbars}, \alpha_{\MSbars}\right)
\right. \nonumber \\
&& \left. ~ -~ \alpha_{\MSbars} \gamma^{\MSbars}_\alpha
\left(a_{\MSbars},\alpha_{\MSbars}\right) 
\frac{\partial ~~~}{\partial \alpha_{\MSbars}}
\ln C_{{\cal O}^T}\left(a_{\MSbars}, \alpha_{\MSbars}\right)
\right]_{\MSbars \to \RIs} ~.~~~~~~
\label{confunrge}
\end{eqnarray}
We have labelled the variables in the two different schemes explicitly for 
clarity. As the right hand side of (\ref{confunrge}) involves variables in the
$\MSbar$ scheme these have to be mapped to their $\RI$ counterparts which is
the meaning of the restriction on the right square bracket. It is a simple 
exercise to infer $\alpha_{\MSbars}\left(a_{\RIs},\alpha_{\RIs}\right)$ from
the three loop expression of 
$\alpha_{\RIs}\left(a_{\MSbars},\alpha_{\MSbars}\right)$ given in \cite{7} to
facilitate this. We have recorded the {\em four} loop Landau gauge expression 
for $C_{{\cal{O}}^T}(a,\alpha)$ for a general colour group in Appendix A with
the full arbitrary gauge expression given in the attached data file. While the
four loop term is not needed to carry out the check of (\ref{gammaTri}) given 
(\ref{gammaTms}) it will in fact be useful once the five loop $\MSbar$ 
expression for (\ref{gammaTms}) is available. As a point of reference we note 
\begin{eqnarray}
\left. \frac{}{} C_{{\cal{O}}^T}(a,0) \right|^{SU(3)} &=&
1 + [ 46.665355 - 3.864197 \Nf ] a^2 \nonumber \\
&& +~ [ 6.763867 \Nf^2 - 308.983059 \Nf + 2060.637793 ] a^3  \nonumber \\
&& +~ [ 97451.822851 - 23266.484197 \Nf + 1309.625138 \Nf^2
\nonumber \\
&& ~~~~
-~ 15.567877 \Nf^3 ] a^4 ~+~ O(a^5)
\end{eqnarray}
numerically. Finally we record that using (\ref{confunrge}) we reproduced
(\ref{gammaTri}) precisely for an arbitrary colour group and gauge parameter.

\sect{Discussion.}

We have evaluated the Green's function with the quark bilinear tensor operator
inserted in a quark $2$-point function to four loops in QCD for both the
$\MSbar$ and the lattice motivated $\RI$ renormalization schemes. As a 
corollary we have deduced the tensor operator anomalous dimension in the 
$\MSbar$ scheme for an arbitrary colour group. To gain an insight into the 
effect of the new loop order we have shown that for $SU(3)$ and three quark 
flavours the four loop correction of the Green's function in the Landau gauge 
represents a insignificantly small difference to the three loop value at the 
same benchmark point in the $\MSbar$ scheme. While this is not unrelated to the
fact that the one loop correction for this Green's function is zero in the 
Landau gauge it perhaps indicates that any error on the lattice extrapolation 
to the high energy continuum value could be very well under control for this 
particular operator. One obvious test of this would be for a re-examination of 
previous lattice extrapolations to the high energy limit but using the new four
loop perturbative results rather than the previous three loop ones.

\vspace{1cm}
\noindent
{\bf Acknowledgements.} This work was carried out with the support of the STFC
Consolidated Grant ST/T000988/1 and partly with the support of a DFG Mercator 
Fellowship. For the purpose of open access, the author has applied a Creative 
Commons Attribution (CC-BY) licence to any Author Accepted Manuscript version 
arising. The data representing the main results here are accessible in 
electronic form from the arXiv ancillary directory associated with the article.

\appendix

\sect{Results for general colour group.}

As the four loop expressions for an arbitrary colour group are large we record 
them here for completeness. First, the $\MSbar$ non-zero amplitude is 
\begin{eqnarray}
\Sigma^{(1) ~ \MSbars}_{{\cal{O}}^T}(p)
&=& 1 
+ \left[
\left[
11 \zeta_3
- \frac{3773}{216}
+ 3 \alpha
- 3 \zeta_3 \alpha
+ \frac{3}{8} \alpha^2
\right]
C_F C_A 
+ \left[
\frac{65}{3}
- 20 \zeta_3
- \alpha^2
\right]
C_F^2 
\right. \nonumber \\
&& \left. ~~~~~~
+ \frac{62}{27}
\Nf T_F C_F 
\right] a^2
\nonumber \\
&&
+ \left[ 
\left[
\frac{251}{16} \zeta_4
- \frac{4180535}{11664}
- \frac{185}{12} \zeta_5
+ \frac{6742}{27} \zeta_3
+ \frac{12817}{576} \alpha
+ \frac{35}{6} \zeta_5 \alpha
+ \frac{3}{8} \zeta_4 \alpha
\right. \right. \nonumber \\
&& \left. \left. ~~~~
- \frac{253}{12} \zeta_3 \alpha
+ \frac{197}{64} \alpha^2
+ \frac{5}{4} \zeta_5 \alpha^2
+ \frac{3}{16} \zeta_4 \alpha^2
- \frac{15}{4} \zeta_3 \alpha^2
+ \frac{29}{48} \alpha^3
- \frac{1}{3} \zeta_3 \alpha^3
\right]
C_F C_A^2 
\right. \nonumber \\
&& \left. ~~~
+ \left[
\frac{62018}{81}
+ 40 \zeta_5
- 50 \zeta_4
- \frac{1862}{3} \zeta_3
- \frac{1}{6} \alpha
+ 20 \alpha \zeta_5
- \frac{79}{3} \zeta_3 \alpha
- 6 \alpha^2
+ 2 \zeta_3 \alpha^2
\right. \right. \nonumber \\
&& \left. \left. ~~~~~~~
- \frac{3}{2} \alpha^3
+ \zeta_3 \alpha^3
\right]
C_F^2 C_A 
\right. \nonumber \\
&& \left. ~~~
+ \left[
32 \zeta_4
- \frac{5246}{27}
- \frac{40}{3} \zeta_5
+ \frac{1550}{9} \zeta_3
+ \alpha
+ 2 \zeta_3 \alpha
- 2 \alpha^2
+ 2 \zeta_3 \alpha^2
- \frac{2}{3} \zeta_3 \alpha^3
\right]
C_F^3 
\right. \nonumber \\
&& \left. ~~~
+ \left[
\frac{79544}{729}
+ 8 \zeta_4
- \frac{1732}{27} \zeta_3
- \frac{673}{72} \alpha
+ 4 \zeta_3 \alpha
\right]
\Nf T_F C_F C_A 
\right. \nonumber \\
&& \left. ~~~
+ \left[
112 \zeta_3
- \frac{23831}{162}
- 8 \zeta_4
+ \frac{4}{3} \alpha
+ \frac{8}{3} \zeta_3 \alpha
\right]
\Nf T_F C_F^2 
\right. \nonumber \\
&& \left. ~~~
+ \left[
\frac{376}{729}
- \frac{32}{27} \zeta_3
\right]
\Nf^2 T_F^2 C_F 
\right] a^3
\nonumber \\
&&
+ \left[ 
\left[
\frac{1568515}{288} \zeta_5
- \frac{3403}{18}
- \frac{282737}{48} \zeta_7
+ \frac{12775}{192} \zeta_6
+ \frac{1675}{64} \zeta_4
- \frac{119405}{144} \zeta_3
\right. \right. \nonumber \\
&& \left. \left. ~~~~
+ \frac{136595}{96} \zeta_3^2
+ \frac{23}{6} \alpha
+ \frac{1939}{96} \zeta_7 \alpha
- \frac{275}{8} \zeta_6 \alpha
- \frac{1415}{6} \zeta_5 \alpha
+ \frac{303}{16} \zeta_4 \alpha
+ \frac{429}{2} \zeta_3 \alpha
\right. \right. \nonumber \\
&& \left. \left. ~~~~
+ \frac{1}{4} \zeta_3^2 \alpha
+ \frac{259}{8} \zeta_7 \alpha^2
- \frac{175}{32} \zeta_6 \alpha^2
- \frac{475}{16} \zeta_5 \alpha^2
+ \frac{33}{32} \zeta_4 \alpha^2
- \frac{65}{24} \zeta_3 \alpha^2
+ \frac{53}{16} \zeta_3^2 \alpha^2
\right. \right. \nonumber \\
&& \left. \left. ~~~~
+ \frac{441}{32} \zeta_7 \alpha^3
- \frac{15}{2} \zeta_5 \alpha^3
- \frac{9}{16} \zeta_4 \alpha^3
- \frac{5}{2} \zeta_3 \alpha^3
- 3 \zeta_3^2 \alpha^3
+ \frac{25}{64} \zeta_6 \alpha^4
+ \frac{75}{32} \zeta_5 \alpha^4
\right. \right. \nonumber \\
&& \left. \left. ~~~~
- \frac{21}{64} \zeta_4 \alpha^4
- \frac{21}{16} \zeta_3 \alpha^4
- \frac{19}{32} \zeta_3^2 \alpha^4
\right]
\frac{d_F^{abcd} d_A^{abcd}}{\NF} 
\right. \nonumber \\
&& \left. ~~~
+ \left[
\frac{69475}{2304} \zeta_6
- \frac{2408522491}{279936}
- \frac{2520203}{2304} \zeta_7
+ \frac{1622471}{1152} \zeta_5
+ \frac{638629}{4608} \zeta_4
\right. \right. \nonumber \\
&& \left. \left. ~~~~~~~
+ \frac{30933595}{5184} \zeta_3
+ \frac{96271}{576} \zeta_3^2
+ \frac{232652657}{746496} \alpha
- \frac{18067}{576} \zeta_7 \alpha
- \frac{2875}{192} \zeta_6 \alpha
\right. \right. \nonumber \\
&& \left. \left. ~~~~~~~
+ \frac{217739}{1728} \zeta_5 \alpha
+ \frac{2727}{256} \zeta_4 \alpha
- \frac{173347}{432} \zeta_3 \alpha
+ \frac{1999}{32} \zeta_3^2 \alpha
+ \frac{1028105}{27648} \alpha^2
\right. \right. \nonumber \\
&& \left. \left. ~~~~~~~
+ \frac{847}{768} \zeta_7 \alpha^2
- \frac{475}{384} \zeta_6 \alpha^2
+ \frac{2309}{144} \zeta_5 \alpha^2
+ \frac{567}{256} \zeta_4 \alpha^2
- \frac{49489}{1152} \zeta_3 \alpha^2
- \frac{197}{96} \zeta_3^2 \alpha^2
\right. \right. \nonumber \\
&& \left. \left. ~~~~~~~
+ \frac{13777}{1536} \alpha^3
+ \frac{147}{128} \zeta_7 \alpha^3
+ \frac{73}{64} \zeta_5 \alpha^3
+ \frac{63}{256} \zeta_4 \alpha^3
- \frac{97}{12} \zeta_3 \alpha^3
- \frac{1}{16} \zeta_3^2 \alpha^3
\right. \right. \nonumber \\
&& \left. \left. ~~~~~~~
+ \frac{529}{384} \alpha^4
+ \frac{25}{768} \zeta_6 \alpha^4
+ \frac{5}{128} \zeta_5 \alpha^4
+ \frac{1}{512} \zeta_4 \alpha^4
- \frac{199}{384} \zeta_3 \alpha^4
\right. \right. \nonumber \\
&& \left. \left. ~~~~~~~
- \frac{5}{192}  \zeta_3^2\alpha^4
\right]
C_F C_A^3 
\right. \nonumber \\
&& \left. ~~~
+ 
\left[
\frac{2009482489}{93312}
+ \frac{521633}{96} \zeta_7
+ \frac{175}{24} \zeta_6
- \frac{25435}{18} \zeta_5
- \frac{23389}{48} \zeta_4
- \frac{4088551}{216} \zeta_3
\right. \right. \nonumber \\
&& \left. \left. ~~~~~~~
- \frac{15779}{12} \zeta_3^2
+ \frac{27239}{432} \alpha
- \frac{791}{12} \zeta_7 \alpha
+ \frac{25}{8} \zeta_6 \alpha
+ \frac{55}{9} \zeta_5 \alpha
- \frac{21}{16} \zeta_4 \alpha
- \frac{25519}{216} \zeta_3 \alpha
\right. \right. \nonumber \\
&& \left. \left. ~~~~~~~
- \frac{103}{4} \zeta_3^2 \alpha
- \frac{42625}{576} \alpha^2
- \frac{427}{32} \zeta_7 \alpha^2
+ \frac{85}{4} \zeta_5 \alpha^2
+ \frac{1315}{72} \zeta_3 \alpha^2
+ \frac{9}{2} \zeta_3^2 \alpha^2
\right. \right. \nonumber \\
&& \left. \left. ~~~~~~~
- \frac{313}{16} \alpha^3
+ \frac{57}{8} \zeta_3 \alpha^3
- \frac{495}{128} \alpha^4
+ \frac{5}{4} \zeta_3 \alpha^4
\right]
C_F^2 C_A^2 
\right. \nonumber \\
&& \left. ~~~
+ 
\left[
\frac{350}{3} \zeta_6
- \frac{224041}{24}
- \frac{71113}{6} \zeta_7
+ \frac{42629}{9} \zeta_5
+ \frac{1655}{6} \zeta_4
+ \frac{89683}{9} \zeta_3
+ \frac{7100}{3} \zeta_3^2
\right. \right. \nonumber \\
&& \left. \left. ~~~~~~~
- \frac{313}{3} \alpha
- \frac{105}{2} \zeta_7 \alpha
+ \frac{530}{3} \zeta_5 \alpha
- \frac{20}{9} \zeta_3 \alpha
+ 20 \zeta_3^2 \alpha
- \frac{3533}{108} \alpha^2
+ \frac{505}{18} \zeta_3 \alpha^2
\right. \right. \nonumber \\
&& \left. \left. ~~~~~~~
- 5 \alpha^3
+ 4 \zeta_3 \alpha^3
+ \frac{5}{8} \alpha^4
- \frac{1}{2} \zeta_3 \alpha^4
\right]
C_F^3 C_A 
\right. \nonumber \\
&& \left. ~~~
+ 
\left[
\frac{57551}{36}
+ \frac{19726}{3} \zeta_7
- \frac{800}{3} \zeta_6
- \frac{44710}{9} \zeta_5
+ 100 \zeta_4
- \frac{10856}{9} \zeta_3
- \frac{2144}{3} \zeta_3^2
\right. \right. \nonumber \\
&& \left. \left. ~~~~~~~
+ 41 \alpha
- 44 \zeta_3 \alpha
- \frac{8}{3} \alpha^2
+ 4 \zeta_3 \alpha^2
\right]
C_F^4 
\right. \nonumber \\
&& \left. ~~~
+ 
\left[
1764 \zeta_7
- \frac{392}{3}
- 1920 \zeta_5
+ 12 \zeta_4
+ \frac{4936}{3} \zeta_3
- 1216 \zeta_3^2
\right]
\Nf \frac{d_F^{abcd} d_F^{abcd}}{\NF} 
\right. \nonumber \\
&& \left. ~~~
+ \left[
\frac{59341063}{11664}
- \frac{735}{4} \zeta_7
- \frac{400}{3} \zeta_6
- \frac{295943}{216} \zeta_5
+ \frac{1951}{16} \zeta_4
- \frac{435797}{216} \zeta_3
- 236 \zeta_3^2
\right. \right. \nonumber \\
&& \left. \left. ~~~~~~~
- \frac{18122533}{93312} \alpha
+ \frac{4081}{54} \zeta_5 \alpha
- \frac{111}{8} \zeta_4 \alpha
+ \frac{656}{9} \zeta_3 \alpha
+ \frac{50}{3} \zeta_3^2 \alpha
- \frac{16801}{1728} \alpha^2
\right. \right. \nonumber \\
&& \left. \left. ~~~~~~~
- \frac{29}{72} \zeta_5 \alpha^2
- \frac{7}{16} \zeta_4 \alpha^2
+ \frac{337}{72} \zeta_3 \alpha^2
+ \zeta_3^2 \alpha^2
\right]
\Nf T_F C_F C_A^2 
\right. \nonumber \\
&& \left. ~~~
+ \left[
882 \zeta_7
- \frac{107169125}{11664}
+ \frac{250}{3} \zeta_6
+ \frac{9674}{9} \zeta_5
- \frac{302}{3} \zeta_4
+ \frac{139813}{27} \zeta_3
+ 868 \zeta_3^2
\right. \right. \nonumber \\
&& \left. \left. ~~~~~~~
- \frac{99637}{864} \alpha
- \frac{494}{9} \zeta_5 \alpha
+ \frac{33}{2} \zeta_4 \alpha
+ \frac{6905}{27} \zeta_3 \alpha
- 32 \zeta_3^2 \alpha
+ \frac{965}{36} \alpha^2
\right. \right. \nonumber \\
&& \left. \left. ~~~~~~~
- \frac{112}{9} \zeta_3 \alpha^2
\right]
\Nf T_F C_F^2 C_A 
\right. \nonumber \\
&& \left. ~~~
+ \left[
\frac{67265}{72}
+ 100 \zeta_6
+ \frac{1624}{9} \zeta_5
- \frac{71}{3} \zeta_4
- \frac{3322}{9} \zeta_3
- 792 \zeta_3^2
+ \frac{476}{9} \alpha
- \frac{488}{9} \zeta_3 \alpha
\right. \right. \nonumber \\
&& \left. \left. ~~~~~~~
+ \frac{94}{27} \alpha^2
- \frac{16}{9} \zeta_3 \alpha^2
\right]
\Nf T_F C_F^3 
\right. \nonumber \\
&& \left. ~~~
+ \left[
\frac{6328}{27} \zeta_5
- \frac{3806959}{5832}
- \frac{56}{3} \zeta_4
+ \frac{580}{3} \zeta_3
+ \frac{35311}{1458} \alpha
- \frac{80}{3} \zeta_5 \alpha
+ 2 \zeta_4 \alpha
\right. \right. \nonumber \\
&& \left. \left. ~~~~~~~
+ \frac{76}{27} \zeta_3 \alpha
\right]
\Nf^2 T_F^2 C_F C_A 
\right. \nonumber \\
&& \left. ~~~
+ \left[
\frac{1291207}{1458}
- \frac{752}{3} \zeta_5
+ \frac{56}{3} \zeta_4
- \frac{4552}{9} \zeta_3
- \frac{208}{27} \alpha
- \frac{128}{27} \zeta_3 \alpha
\right]
\Nf^2 T_F^2 C_F^2 
\right. \nonumber \\
&& \left. ~~~
+ \left[
\frac{9220}{2187}
- \frac{16}{9} \zeta_4
+ \frac{64}{81} \zeta_3
\right]
\Nf^3 T_F^3 C_F 
\right] a^4 ~+~ O(a^5) ~.
\end{eqnarray}
The conversion function for the tensor operator from the $\MSbar$ scheme to the
$\RI$ one is 
\begin{eqnarray}
C_{{\cal{O}}^T}(a,0) &=& 1 + \left[ 5987 C_A - 3024 \zeta_3 C_A 
+ 4320 \zeta_3 C_F - 4815 C_F - 1252 \Nf T_F \right] \frac{C_A a^2}{216} 
\nonumber \\
&&
+ \left[ 
660960 \zeta_5 C_A^2
- 233280 \zeta_4 C_A^2
- 4438098 \zeta_3 C_A^2
+ 7047161 C_A^2 
+ 7752672 \zeta_3 C_A C_F
\right. \nonumber \\
&& \left. ~~~
+ 653184 \zeta_4 C_A C_F
- 699840 \zeta_5 C_A C_F
- 9415134 C_A C_F 
+ 950400 \zeta_3 C_A \Nf T_F
\right. \nonumber \\
&& \left. ~~~
- 93312 \zeta_4 C_A \Nf T_F
- 2984432 C_A \Nf T_F 
- 2008800 \zeta_3 C_F^2
- 373248 \zeta_4 C_F^2
\right. \nonumber \\
&& \left. ~~~
+ 155520 \zeta_5 C_F^2
+ 2195316 C_F^2 
- 1119744 \zeta_3 C_F \Nf T_F
+ 93312 \zeta_4 C_F \Nf T_F
\right. \nonumber \\
&& \left. ~~~
+ 1562256 C_F \Nf T_F 
+ 13824 \zeta_3 \Nf^2 T_F^2
+ 220064 \Nf^2 T_F^2 \right] \frac{C_F a^3}{11664} \nonumber \\
&&
+ \left[
859248 \zeta_3^2 C_A^3 C_F
- 5427947484 \zeta_3 C_A^3 C_F
- 97962048 \zeta_4 C_A^3 C_F
\right. \nonumber \\
&& \left. ~~~
+ 307910160 \zeta_5 C_A^3 C_F
- 62208000 \zeta_6 C_A^3 C_F 
+ 226816443 \zeta_7 C_A^3 C_F
\right. \nonumber \\
&& \left. ~~~
+ 7769141516 C_A^3 C_F 
+ 773276544 \zeta_3^2 C_A^2 C_F^2
+ 9867578112 \zeta_3 C_A^2 C_F^2
\right. \nonumber \\
&& \left. ~~~
+ 228614400 \zeta_4 C_A^2 C_F^2
+ 907933536 \zeta_5 C_A^2 C_F^2
+ 270604800 \zeta_6 C_A^2 C_F^2
\right. \nonumber \\
&& \left. ~~~
- 2066061816 \zeta_7 C_A^2 C_F^2
- 12614153610 C_A^2 C_F^2 
+ 179159040 \zeta_3^2 C_A^2 C_F \Nf T_F
\right. \nonumber \\
&& \left. ~~~
+ 2217673728 \zeta_3 C_A^2 C_F \Nf T_F
- 68584320 \zeta_4 C_A^2 C_F \Nf T_F
\right. \nonumber \\
&& \left. ~~~
+ 271216512 \zeta_5 C_A^2 C_F \Nf T_F
+ 46656000 \zeta_6 C_A^2 C_F \Nf T_F
\right. \nonumber \\
&& \left. ~~~
- 20575296 \zeta_7 C_A^2 C_F \Nf T_F
- 5293901856 C_A^2 C_F \Nf T_F 
- 1517626368 \zeta_3^2 C_A C_F^3
\right. \nonumber \\
&& \left. ~~~
- 2501316288 \zeta_3 C_A C_F^3
+ 23607936 \zeta_4 C_A C_F^3
- 2880479232 \zeta_5 C_A C_F^3
\right. \nonumber \\
&& \left. ~~~
- 569203200 \zeta_6 C_A C_F^3
+ 3784548096 \zeta_7 C_A C_F^3
+ 5214521988 C_A C_F^3 
\right. \nonumber \\
&& \left. ~~~
- 575548416 \zeta_3^2 C_A C_F^2 \Nf T_F
- 2772776448 \zeta_3 C_A C_F^2 \Nf T_F
\right. \nonumber \\
&& \left. ~~~
+ 56360448 \zeta_4 C_A C_F^2 \Nf T_F
- 535735296 \zeta_5 C_A C_F^2 \Nf T_F
\right. \nonumber \\
&& \left. ~~~
+ 9331200 \zeta_6 C_A C_F^2 \Nf T_F
+ 4472230512 C_A C_F^2 \Nf T_F 
\right. \nonumber \\
&& \left. ~~~
- 160496640 \zeta_3 C_A C_F \Nf^2 T_F^2
+ 17169408 \zeta_4 C_A C_F \Nf^2 T_F^2
\right. \nonumber \\
&& \left. ~~~
- 103845888 \zeta_5 C_A C_F \Nf^2 T_F^2
+ 1012343136 C_A C_F \Nf^2 T_F^2 
\right. \nonumber \\
&& \left. ~~~
+ 579280896 \zeta_3^2 C_F^4
- 905561856 \zeta_3 C_F^4
- 139968000 \zeta_4 C_F^4
\right. \nonumber \\
&& \left. ~~~
+ 2238243840 \zeta_5 C_F^4
+ 373248000 \zeta_6 C_F^4
- 812464182 C_F^4 
\right. \nonumber \\
&& \left. ~~~
+ 443418624 \zeta_3^2 C_F^3 \Nf T_F
+ 58371840 \zeta_3 C_F^3 \Nf T_F
- 186624 \zeta_4 C_F^3 \Nf T_F
\right. \nonumber \\
&& \left. ~~~
- 97293312 \zeta_5 C_F^3 \Nf T_F
- 55987200 \zeta_6 C_F^3 \Nf T_F
- 339420240 C_F^3 \Nf T_F 
\right. \nonumber \\
&& \left. ~~~
+ 134618112 \zeta_3 C_F^2 \Nf^2 T_F^2
- 17169408 \zeta_4 C_F^2 \Nf^2 T_F^2
+ 140341248 \zeta_5 C_F^2 \Nf^2 T_F^2
\right. \nonumber \\
&& \left. ~~~
- 282227232 C_F^2 \Nf^2 T_F^2 
- 1437696 \zeta_3 C_F \Nf^3 T_F^3
+ 995328 \zeta_4 C_F \Nf^3 T_F^3
\right. \nonumber \\
&& \left. ~~~
- 51645184 C_F \Nf^3 T_F^3 
- 710384256 \zeta_3^2 \frac{d_F^{abcd} d_A^{abcd}}{\NF}
+ 485696736 \zeta_3 \frac{d_F^{abcd} d_A^{abcd}}{\NF}
\right. \nonumber \\
&& \left. ~~~
+ 14556672 \zeta_4 \frac{d_F^{abcd} d_A^{abcd}}{\NF}
- 2831980320 \zeta_5 \frac{d_F^{abcd} d_A^{abcd}}{\NF}
- 74649600 \zeta_6 \frac{d_F^{abcd} d_A^{abcd}}{\NF}
\right. \nonumber \\
&& \left. ~~~
+ 2969129520 \zeta_7 \frac{d_F^{abcd} d_A^{abcd}}{\NF}
- 21523968 \frac{d_F^{abcd} d_A^{abcd}}{\NF} 
+ 465813504 \zeta_3^2 \Nf \frac{d_F^{abcd} d_F^{abcd}}{\NF}
\right. \nonumber \\
&& \left. ~~~
- 291879936 \zeta_3 \Nf \frac{d_F^{abcd} d_F^{abcd}}{\NF}
- 6718464 \zeta_4 \Nf \frac{d_F^{abcd} d_F^{abcd}}{\NF}
\right. \nonumber \\
&& \left. ~~~
+ 380712960 \zeta_5 \Nf \frac{d_F^{abcd} d_F^{abcd}}{\NF}
- 987614208 \zeta_7 \Nf \frac{d_F^{abcd} d_F^{abcd}}{\NF}
\right. \nonumber \\
&& \left. ~~~
+ 338535936 \Nf \frac{d_F^{abcd} d_F^{abcd}}{\NF} \right]
\frac{a^4}{559872} ~+~ O(a^5)
\end{eqnarray}
where $\NF$ is the dimension of the fundamental representation. Finally the
tensor operator anomalous dimension in the $\RI$ scheme is 
\begin{eqnarray}
\gamma_{{\cal{O}}^T}^{\RIs}(a,\alpha) &=&
C_F a 
+ \left[ 9 \alpha^2 C_A 
+ 27 \alpha C_A 
+ 257 C_A 
- 171 C_F 
- 52 \Nf T_F \right] \frac{C_F a^2}{18} \nonumber \\
&& 
+ \left[ 162 \alpha^4 C_A^2 
+ 1215 \alpha^3 C_A^2 
+ 324 \alpha^3 C_A C_F 
+ 5715 \alpha^2 C_A^2 
+ 972 \alpha^2 C_A C_F 
\right. \nonumber \\
&& \left. ~~~
- 1440 \alpha^2 C_A \Nf T_F 
+ 16902 \alpha C_A^2 
- 6264 \alpha C_A \Nf T_F 
- 92448 \zeta_3 C_A^2 
\right. \nonumber \\
&& \left. ~~~
+ 213548 C_A^2 
+ 167616 \zeta_3 C_A C_F 
- 228744 C_A C_F 
+ 13824 \zeta_3 C_A \Nf T_F 
\right. \nonumber \\
&& \left. ~~~
- 99536 C_A \Nf T_F 
- 41472 \zeta_3 C_F^2 
+ 39420 C_F^2 
- 24192 \zeta_3 C_F \Nf T_F 
\right. \nonumber \\
&& \left. ~~~
+ 45576 C_F \Nf T_F 
+ 9152 \Nf^2 T_F^2 \right] \frac{C_F a^3}{648} \nonumber \\
&& 
+ \left[ 3645 \alpha^6 C_A^3 C_F 
+ 40824 \alpha^5 C_A^3 C_F 
+ 8748 \alpha^5 C_A^2 C_F^2 
+ 272403 \alpha^4 C_A^3 C_F 
\right. \nonumber \\
&& \left. ~~~
+ 49572 \alpha^4 C_A^2 C_F^2 
- 38880 \alpha^4 C_A^2 C_F \Nf T_F 
+ 11664 \alpha^4 C_A C_F^3 
+ 24786 \zeta_3 \alpha^3 C_A^3 C_F 
\right. \nonumber \\
&& \left. ~~~
+ 1104192 \alpha^3 C_A^3 C_F 
+ 23328 \zeta_3 \alpha^3 C_A^2 C_F^2 
+ 210924 \alpha^3 C_A^2 C_F^2 
\right. \nonumber \\
&& \left. ~~~
- 309096 \alpha^3 C_A^2 C_F \Nf T_F 
- 46656 \zeta_3 \alpha^3 C_A C_F^3 
+ 81648 \alpha^3 C_A C_F^3 
\right. \nonumber \\
&& \left. ~~~
- 77760 \alpha^3 C_A C_F^2 \Nf T_F 
- 319788 \zeta_3 \alpha^2 C_A^3 C_F 
- 38880 \zeta_5 \alpha^2 C_A^3 C_F 
\right. \nonumber \\
&& \left. ~~~
+ 4203198 \alpha^2 C_A^3 C_F 
+ 299376 \zeta_3 \alpha^2 C_A^2 C_F^2 
+ 365472 \alpha^2 C_A^2 C_F^2 
\right. \nonumber \\
&& \left. ~~~
- 104976 \zeta_3 \alpha^2 C_A^2 C_F \Nf T_F 
- 2033856 \alpha^2 C_A^2 C_F \Nf T_F 
- 334368 \zeta_3 \alpha^2 C_A C_F^3 
\right. \nonumber \\
&& \left. ~~~
+ 316872 \alpha^2 C_A C_F^3 
+ 311040 \zeta_3 \alpha^2 C_A C_F^2 \Nf T_F 
- 645408 \alpha^2 C_A C_F^2 \Nf T_F 
\right. \nonumber \\
&& \left. ~~~
+ 172800 \alpha^2 C_A C_F \Nf^2 T_F^2 
+ 62208 \zeta_3 \alpha^2 C_F^3 \Nf T_F 
- 62208 \alpha^2 C_F^3 \Nf T_F 
\right. \nonumber \\
&& \left. ~~~
- 2959578 \zeta_3 \alpha C_A^3 C_F 
- 686880 \zeta_5 \alpha C_A^3 C_F 
+ 13835772 \alpha C_A^3 C_F 
\right. \nonumber \\
&& \left. ~~~
+ 2541456 \zeta_3 \alpha C_A^2 C_F^2 
- 1253556 \alpha C_A^2 C_F^2 
+ 990144 \zeta_3 \alpha C_A^2 C_F \Nf T_F 
\right. \nonumber \\
&& \left. ~~~
+ 207360 \zeta_5 \alpha C_A^2 C_F \Nf T_F 
- 9117360 \alpha C_A^2 C_F \Nf T_F 
- 412128 \zeta_3 \alpha C_A C_F^3 
\right. \nonumber \\
&& \left. ~~~
+ 103032 \alpha C_A C_F^3 
- 196992 \zeta_3 \alpha C_A C_F^2 \Nf T_F 
- 977184 \alpha C_A C_F^2 \Nf T_F 
\right. \nonumber \\
&& \left. ~~~
- 248832 \zeta_3 \alpha C_A C_F \Nf^2 T_F^2 
+ 1347840 \alpha C_A C_F \Nf^2 T_F^2 
+ 124416 \zeta_3 \alpha C_F^3 \Nf T_F 
\right. \nonumber \\
&& \left. ~~~
- 31104 \alpha C_F^3 \Nf T_F 
+ 165888 \zeta_3 \alpha C_F^2 \Nf^2 T_F^2 
- 41472 \alpha C_F^2 \Nf^2 T_F^2 
\right. \nonumber \\
&& \left. ~~~
- 115925580 \zeta_3 C_A^3 C_F 
+ 10393920 \zeta_5 C_A^3 C_F 
+ 195274634 C_A^3 C_F 
\right. \nonumber \\
&& \left. ~~~
+ 206535744 \zeta_3 C_A^2 C_F^2 
+ 2643840 \zeta_5 C_A^2 C_F^2 
- 271766556 C_A^2 C_F^2 
\right. \nonumber \\
&& \left. ~~~
+ 53147664 \zeta_3 C_A^2 C_F \Nf T_F 
- 2177280 \zeta_5 C_A^2 C_F \Nf T_F 
- 144291864 C_A^2 C_F \Nf T_F 
\right. \nonumber \\
&& \left. ~~~
- 41570496 \zeta_3 C_A C_F^3 
- 34525440 \zeta_5 C_A C_F^3 
+ 87039360 C_A C_F^3 
\right. \nonumber \\
&& \left. ~~~
- 84001536 \zeta_3 C_A C_F^2 \Nf T_F 
+ 6220800 \zeta_5 C_A C_F^2 \Nf T_F 
+ 115518384 C_A C_F^2 \Nf T_F 
\right. \nonumber \\
&& \left. ~~~
- 5391360 \zeta_3 C_A C_F \Nf^2 T_F^2 
+ 30575616 C_A C_F \Nf^2 T_F^2 
- 15552000 \zeta_3 C_F^4 
\right. \nonumber \\
&& \left. ~~~
+ 24883200 \zeta_5 C_F^4 
- 10195308 C_F^4 
+ 12317184 \zeta_3 C_F^3 \Nf T_F 
\right. \nonumber \\
&& \left. ~~~
- 4976640 \zeta_5 C_F^3 \Nf T_F 
- 10896768 C_F^3 \Nf T_F 
+ 7050240 \zeta_3 C_F^2 \Nf^2 T_F^2 
\right. \nonumber \\
&& \left. ~~~
- 10107648 C_F^2 \Nf^2 T_F^2 
- 1744384 C_F \Nf^3 T_F^3 
+ 1617408 \zeta_3 \frac{d_F^{abcd} d_A^{abcd}}{\NF} 
\right. \nonumber \\
&& \left. ~~~
- 4976640 \zeta_5 \frac{d_F^{abcd} d_A^{abcd}}{\NF} 
- 248832  \frac{d_F^{abcd} d_A^{abcd}}{\NF}
- 746496 \zeta_3 \Nf \frac{d_F^{abcd} d_F^{abcd}}{\NF} 
\right. \nonumber \\
&& \left. ~~~
+ 2985984 \Nf \frac{d_F^{abcd} d_F^{abcd}}{\NF} 
\right] \frac{a^4}{23328} ~+~ O(a^5) 
\end{eqnarray}
for an arbitrary linear covariant gauge. The one loop term is clearly scheme
independent.

\end{document}